\documentstyle[psfig]{lamuphys}
\makeatletter
\let\chapter\hid@chapter
\makeatother
\begin{document}
\pagenumbering{arabic}
\title{The nature and observability of protogalaxies}

\author{Simon D.M.\,White\inst{1}}

\institute{Max-Planck-Institut f\"ur Astrophysik,
Karl-Schwarzschild-Stra{\ss}e 1,\\
D-85740 Garching bei M\"unchen, Germany}

\maketitle

\begin{abstract}
I discuss recent theoretical work on the formation and evolution of
galaxies paying particular attention to the ability of current models
to make detailed compar\-isons with observations of the galaxy population
both nearby and at high redshift. These models suggest that much
(perhaps most) star formation in the universe took place in
objects that are already detected in deep galaxy samples. In addition,
they predict that systems with large star formation rates are unlikely
to be much more abundant in the past than they are at present. Recent
data show that the star formation rate in the nearby
universe is, in fact, a substantial fraction of that required to make
all the stars seen in galaxies, and that the observed abundance of objects
forming stars at rates in excess of 10M$_\odot$/yr is approximately
the same at redshifts of 1.25 and 3.25 as it is at $z=0$. Both the 
epoch of galaxy formation and ``typical'' protogalaxies may already
have been observed but not recognised. Thermal
emission from dust in such protogalaxies could be detected 
by a large millimeter array, and molecular line observations could
explore the dynamical state of the gas in the more massive systems.
\end{abstract}
\section{Introduction}

The traditional approach to interpreting observational data on the
faint galaxy population is grounded firmly in the work which Beatrice
Tinsley carried out in the 1970's (see Tinsley 1980). One starts from
a characterisation of the local galaxy population in terms of the 
abundance of objects as a function of luminosity and Hubble type. One
assumes that star formation always occurs with an Initial Mass
Function (IMF) similar to that inferred from observations of the disk 
population in the Solar Neighborhood. One picks a simple parameterized
model for the star-formation history of a galaxy specified, for
example, by the redshift at which star formation starts, the fraction of
the stars form in an initial burst, and the characteristic timescale
with which formation of the remaining stars decays
exponentially in time. These parameters are then adjusted separately
for each Hubble type so that the synthesised stellar population at
redshift zero has colours which agree with those observed for nearby
galaxies. These simple models then allow one to predict the luminosity
and colour which a present day galaxy would have if it were seen, for
example, at $z=1$.

This simple scheme can make full use of the information we have about
nearby galaxies and of our knowledge of stellar evolution. In
addition, it is easily extended to include, for example, a separate
treatment of the bulges and disks of spiral galaxies, or a
phenomenological treatment of starbursts. It allows one to use
the counts, colours and redshift distributions of faint galaxies, to
look for possible constraints on cosmological parameters and on the
``epoch of galaxy formation''. Furthermore, the assumption of a well
defined collapse epoch when a burst forms a significant fraction of
the final stellar population has been traditional since the earliest
discussions of galaxy formation (e.g. Partridge and Peebles 1967) and
has virtue of predicting that the collapsing protogalaxy should be
very bright and so, perhaps, easily observable. The failure to find
such objects has convinced many observers that galaxy formation must 
occur at high redshift. An interesting 
variant of the traditional approach is presented by Gronwall and
Koo (1995) who show that it can still explain much of
the recent data.

With the refurbishment of the HST and the advent of spectroscopy on
10 meter telescopes, this kind of approach is no longer adequate.
Samples of field galaxies can now be studied to redshift 3 and beyond,
and the new morphological and spectral data show that these objects
cannot simply be considered as present-day galaxies whose
star-formation is at a less advanced stage but whose structure is
otherwise unaltered. The distant objects are often disturbed, are
in most cases relatively small, and typically show evidence for
substantial ongoing star formation (Cowie et al 1995; Steidel et
al 1996; Giavalisco et al 1996, Abraham et al 1996). It seems likely 
that an understanding of their relation to nearby systems will only 
be obtained through more detailed consideration of how different 
types of galaxies were assembled and made their stars. 

Theoretical understanding of the formation and evolution of galaxies
has improved greatly over the past two decades. Surveys of the 
spatial and kinematic distributions of 
galaxies, together with simulations which demonstrate how 
gravity can produce these distributions, have led to a
much clearer picture of the likely context for galaxy 
formation than was available in the 1970's. This picture can be
further explored at high redshift using quasar absorption lines
(e.g. Cen et al 1994; Katz et al 1996). In addition,
studies of the stellar populations within galaxies and of interacting 
galaxies have highlighted the role of galaxy transformation processes 
like starbursts and mergers. As I now discuss, 
hierarchical clustering theory can be used to a build a phenomenology of
galaxy formation which is physically based, relatively simple, tuned
to agree with detailed simulation results where these are available,
and designed to allow direct comparison with observation.

\section{The current structure formation paradigm}

A well-developed ``standard'' picture for the formation of
structure has been adopted as a working hypothesis
by most cosmologists. Its main elements are the 
following.\vskip0.1truecm
\noindent $\bullet$ Most of the matter in the universe is in some
dark, nonbaryonic form. It must be dark because all attempts
to observe radiation from the dominant component in galaxy
clusters or in the outer halos of galaxies have failed.
It must be nonbaryonic because the mass content of the universe
inferred from dynamical
and gravitational lensing measurements exceeds the baryonic content
required if the observed abundances of light elements is to be
consistent 
with the theory of Big Bang nucleosynthesis.\vskip0.1truecm
\noindent $\bullet$ Baryonic matter is present in the amount predicted by Big
Bang nucleosynthesis. This is few percent of the closure density, and is
significantly larger than the amount observed directly in the form of
stars or intergalactic gas.\vskip0.1truecm
\noindent $\bullet$ Deviations from uniformity were small in the early universe
and were generated at very early times, probably by quantum effects
during inflation. This process imposed no
characteristic scale relevant to the formation of galaxies or larger
structures, and produced a gaussian field of linear density
fluctuations. An alternative, structure generation at late times 
by topological defects such as cosmic strings 
or textures, has attracted less attention because its consequences are
much harder to calculate.\vskip0.1truecm
\noindent $\bullet$ Structure grows through gravitational instability. Radiation
pressure on the gas and dispersive motions in the dark matter can 
have significant effects at early times, but
later evolution is driven entirely by the gravity of
the dark matter until the collapse of 
the dark halos of galaxies.\vskip0.1truecm
\noindent $\bullet$ Galaxies form at late
times by the dissipative settling of gas to the centres of the
potential wells provided by dark matter halos. This explains the
observed segregation between luminous and dark matter in galaxies as
well as the origin of the spin of galactic disks.\hfil\break

At cosmology conferences this framework is usually taken for granted; 
argu\-ments tend to centre on the parameter values which define specific
implement\-ations of it ($H_0, \Omega, \Omega_b, \Lambda$, dark matter
type, etc.). For non-specialists, however, it can seem more relevant
to question whether the universe really is made primarily of some 
entirely new form of matter, and whether all the structure we 
see really arose from quantum zero-point fluctuations when
the universe was $\sim 10^{-35}$s old. As a scheme for forming
galaxies and larger structures, this hierarchical picture was outlined
by White
and Rees (1978) and has been developed in recent years both
by numerical simulations (e.g. Katz et al 1992; 
Navarro et al 1995) and by detailed semi-analytic
treatments of the relevant physical processes.

\section{Hierarchical galaxy formation}

The formation of galaxies is expected to occur in a very similar way
in most currently popular versions of the paradigm just discussed. 
(These are usually designated by acronyms, for example, SCDM, CHDM, 
$\tau$CDM, OCDM, $\Lambda$CDM, TCDM, PIB, texture+HDM; I will not here
discuss the relative merits of these possibilities.)
The dominant processes shaping the evolution of the galaxy popula\-tion
and the structure and morphology of galaxies are the
following.\vskip0.1truecm
\noindent $\bullet$ The dark matter clusters hierarchically from a gaussian
field of initial density fluctuations. Small objects form first and
merge together to make larger ones. A well-developed
analytic theory for the dynamics and statistics of this process and
also for the structure of the resulting objects has been tested
in considerable detail against numerical simulations (Lacey 
\& Cole 1993, 1994; Cole \& Lacey 1996; Mo \& White 1996; Navarro
et al 1996).\vskip0.1truecm
\noindent$\bullet$ Gas cools and collects at the centres of dark halos to form
cold rotationally supported disks. As first shown by Fall and
Efstathiou (1980), the observed angular momenta of spiral galaxy disks
can be produced by tidal torques at early times {\it only} if the
disks formed within an extended massive halo in this
way.\vskip0.1truecm
\noindent $\bullet$ Star formation occurs: (a) in quiescent disks; (b) in bursts
during galaxy collisions and mergers; (c) during the initial collapse
of a galaxy/halo system. In the present universe most star formation 
occurs in mode (a), but mode (b) is also significant. In a
hierarchical model there is little distinction between collapse
and merging, and so between modes (b) and (c).\vskip0.1truecm
\noindent $\bullet$ Winds from massive stars and shocks from supernovae may 
reheat cold gas in galaxy disks and may affect the accretion of new
gas from the surrounding halo. Such feedback effects appear necessary
to limit the efficiency of star formation and to ensure that
sufficient gas remains at late times to form the large galaxies which
contain most observed stars. Direct evidence for
strong feedback associated with vigorous star
formation is seen in the ``superwinds''generated by some
starburst galaxies (Heckman et al 1990).\vskip0.1truecm
\noindent $\bullet$ Ellipticals form by the merger of disk/bulge systems made
primarily of stars.  Gas may
condense to form a new disk around such an elliptical and so transform
it back into the bulge of a spiral. The first process is observed
directly in the nearby universe, although there is controversy over
the fraction of ellipticals produced by it. Direct
evidence for the second can, perhaps, be found in the discovery
of a nearby spiral with a counter-rotating bulge (Prada et al 1996).
\vskip0.1truecm
\noindent $\bullet$ When dark halos merge the galaxies within them
remain distinct. Thus a massive ``cluster'' dark halo can contain
many galaxies, and the halo of a spiral galaxy may include
a number of small satellites. Galaxy mergers occur when
dynamical friction brings the orbit of a subsidiary galaxy near the
centre of its dark halo where it can encounter the dominant galaxy which 
resides there. Thus the Milky Way will accrete the
Magellanic Clouds, and the cD galaxies at cluster centres can grow by
swallowing other cluster galaxies as well as by accreting more gas
from surrounding cooling flows.\hfil\break

All of the above processes seem certain to play a role in
shaping the observed galaxy population. The difficulty lies in
specifying when they should occur and what their relative importance
should be. In addition one needs methods to calculate their effect.
The analytic description of 
hierarchical clustering referred to above can be extended to
generate Monte Carlo realisations of the full merging 
history of a present-day dark halo (Kauffmann \& White 1993). Within
such a merger tree it is possible use simplified analytic descriptions
of the relevant physical processes to follow the formation,
evolution and interaction of all the galaxies which end up in the
final halo. By considering many halos with the appropriate mass
distribution, it is then possible to reconstruct the
galaxy population in a representative region of the universe.

This programme was first carried out by Kauffmann et al
(1993) who showed that if parameters are tuned so that a
dark halo with circular velocity of 220 km/s typically contains a
system resembling the Milky Way, then hierarchical 
models reproduce many of the observed regularities of
the local galaxy population. For example, the Tully-Fisher relation,
the bulge-to-disk ratios of spirals, the morphology-environment and
gas fraction-environment relations, the colour-\break morphology relation,
the rich cluster luminosity function and luminosity-\break morphology relation, 
all these are well reproduced. The most serious discrepancy
affects the field galaxy luminosity function which has the correct
number of bright galaxies but too many faint ones. Models 
which come close to fitting the nearby galaxy
population can also be consistent (with no further adjustment) with
the available counts and redshift distributions for faint galaxies
(Kauffmann et al 1994). An independent implementation
of this programme by Cole et al (1994) differs substantially in many
details but comes to similar conclusions. In Heyl et al (1995) these
authors also studied how the success of such schemes
depends on the particular cosmology in which they are
implemented.

Such attempts to build physically based models differ fundamentally
from the traditional Tinsley approach. Both schemes use
population synthesis models
to predict the photometric properties of galaxies, but the
resemblance stops there. The traditional approach treats each galaxy
independently of all others. The galaxy population at high redshift can
be identified one-to-one with the nearby population whose
properties are adopted from observation rather than derived from a
model. The epoch of galaxy formation is a parameter of
the arbitrary analytic form chosen to describe star formation
histories, and is adjusted to fit faint galaxy data. In contrast, 
galaxies form and transform continually in the hierarchical models.
There is no simple relationship
between the galaxies at $z=1$ and those at $z=0$. Some old ellipticals
have grown new disks, some old spirals have merged to make new
ellipticals, and a significant amount of the matter in nearby galaxies
was in {\it no} galaxy at $z=1$. The parameters which are adjusted 
describe the efficiencies of uncertain physical processes
(for example, star formation, feedback and dynamical friction) and
only indirectly influence the formation history of galaxies. The
present properties of the galaxy population are not built in {\it a
priori}, but are predictions of the model; as a result they usually show
some discrepancies with observation. 

The advantage of the hierarchical models is that because they
offer a complete, albeit schematic, description of the 
history of the galaxy population, they can be used to address a very
broad range of questions. For example, with Monte Carlo realisations
of the full formation history of cluster galaxies one can study the
conditions necessary for an observable Butcher-Oemler effect, for
the elliptical colour-luminosity relation to be as tight as
observed, and for this relation to evolve as observed. One can also
check whether the star formation histories of ``field'' ellipticals
(and thus their  colours) are expected to differ systematically from
those of cluster ellipticals. These questions were addressed by
Kauffmann (1995, 1996) who found that the observed strength of the
Butcher-Oemler effect appears to require a high density universe, that
such a universe can be consistent with the tight colour-luminosity
relation of cluster ellipticals, and that field ellipticals are 
predicted to be younger (and so bluer) than cluster ellipticals.

Figure 1 shows the average formation history predicted for elliptical galaxies in rich clusters for a CDM universe with
$\Omega=1$ and $H_0=50$ km/s/Mpc. About 55\% of the stars in the
ellipticals formed more than 10 Gyr ago and almost none less than 4 Gyrs
ago. On the other hand, the typical elliptical had its last major
merger about 7 Gyrs ago, so that most of the stars were formed well
before the observed galaxies were assembled. From this figure we can
infer that the stars in, say, a $10^{11}$M$_\odot$ elliptical galaxy
were forming rapidly around 11 Gyr ago, corresponding to $z=2.5$.
At this time the star formation rate was of order $0.2\times
10^{11}$M$_\odot/1$Gyr$ = 20$M$_\odot$/yr. However, the galaxy
was in several pieces, so the star formation rate in
the largest piece was $\sim 10$M$_\odot$/yr. This is an order of
magnitude smaller than the rate inferred in traditional models where a
bright elliptical forms in a single burst of duration
1 Gyr or less.

\begin{figure}
\centerline{
\psfig{figure=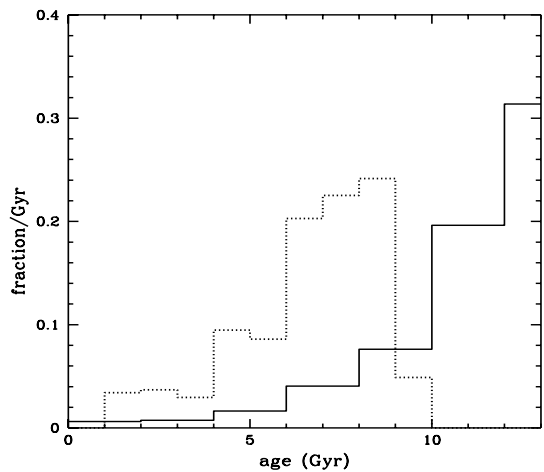,width=7.4cm,height=7.1cm}
}
\caption{The solid histogram is the stellar age distribution for
elliptical galaxies in rich galaxy clusters according to a
standard CDM model with $\Omega=1$ and $H_0=50$ km/s/Mpc. The dotted
histogram gives the time of the last major merger for this same
population of galaxies. For field ellipticals both histograms would be
shifted to lower ages. For field galaxies in general the stellar age
distribution becomes much flatter. (Adapted from fig. 3 of Kauffmann
1996).}
\end{figure}
 
Hierarchical models automatically specify luminosities and
morphologies for all galaxies at all redshifts, and so it
is easy to predict galaxy counts as a function of morphology
as well as of colour and apparent magnitude. In the Hubble Deep Field 
galaxies can be classified to much fainter limits than was previously 
possible. Baugh et al (1996) show that the counts presented 
by Abraham et al (1996) are in good agreement with their previously 
published hierarchical models -- the rapidly rising count of of 
irregular and star-forming systems is reproduced quite naturally. 
The models also predict redshift distributions as a function of
morphology, and it should soon be possible to test these
predictions directly. Hierarchical models can also
predict the spatial distribution of galaxies as a function of
luminosity and type. A first analysis has been presented by
Kauffmann et al (1996) who show that the observed
difference in clustering between spirals and ellipticals and between
bright galaxies and dwarfs is easily reproduced. However, in the rest
of this contribution I concentrate on the star formation history
of the galaxy population since it is this which relates most directly 
to millimeter observations. The case of cluster
ellipticals, discussed above, illustrates the major point; star
formation occurs at lower redshifts, in smaller
objects, and at lower rates than in traditional models. 

\section{The abundance of star-forming galaxies}

The best observational measure of the star-formation rate in nearby 
galaxies is generally thought to be H$\alpha$ luminosity. A recent
survey by Gallego et al (1995) has determined the first reliable 
luminosity function for the local universe based purely on H$\alpha$ 
selection. These authors find that the galaxies which provide most 
of the nearby star-formation differ from those which dominate the optical 
luminosity density. The star-forming galaxies are typic\-ally late-type 
systems with luminosities well below $L_*$ and with star formation 
occurring predominantly in a compact nuclear region. The H$\alpha$
luminosities can be converted to star-formation rates assuming a
standard solar neighborhood IMF, and the data then give the local
abundance of galaxies as a function of star-formation rate. This
distribution is shown in cumulative form in Figure 2. By integrating
over all luminosities Gallego et al estimate a total star-formation rate
of 0.013 M$_\odot$/yr/Mpc$^3$ where here and below $H_0=50$ km/s/Mpc.

It is interesting to compare this number with the total mass density 
of stars in the local universe. The APM redshift sample of
Loveday et al (1992) covers the largest volume surveyed to date and
leads to an estimated luminosity density of $6.5\times
10^7$L$_\odot$/Mpc$^3$ in the B band. According to Efstathiou et al
(1988) 40\% of the B light comes from E/S0 galaxies and the rest from
spirals. The detailed kinematic study of van der Marel (1991) shows
that the mean $M/L$ of early type galaxies is 3.2 in B after
averaging over an appropriate luminosity function. For spirals Persic
and Salucci (1992) estimate $M/L_B\approx 1.0$ after subtracting the
contribution of dark halos to the estimated dynamical masses. Putting
all these numbers together gives a total star density of $1.2\times
10^8$M$_\odot$/Mpc$^3$. Thus the local star formation rate observed by
Gallego et al appears enough to make all the stars observed by Loveday
et al in about 10 Gyrs. The overall star formation rate need never
have been higher in the past than it is today.

In fact there is mounting evidence that the luminosity density
inferred from the APM sample is too low by about a factor of 2,
perhaps because of photometric problems (Ellis et al 1996; Bertin \&
Dennefeld 1996). In this case the average past star formation rate
would need to be twice the currently observed value. This seems
consistent with the detection of strongly enhanced star
formation in intrinsically {\it faint} galaxies at redshifts beyond
0.3 (Ellis et al 1996; Lilly et al 1995). However, these same samples show
little evidence for an enhancement of the abundance of rapidly
star-forming systems, and in fact, as I now show, there appears to 
be no such enhancement out to redshifts beyond 3.

\begin{figure}
\centerline{
\psfig{figure=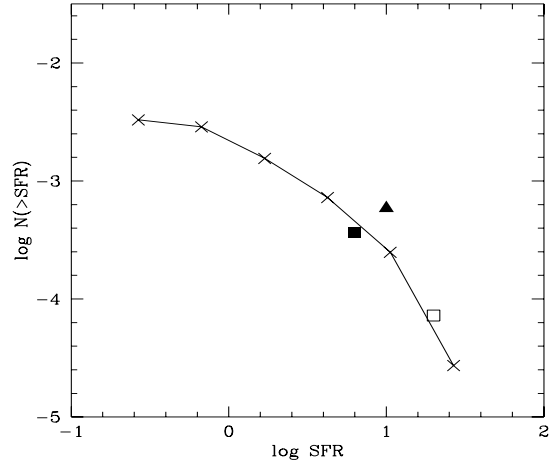,width=7.4cm,height=7.1cm}
}
\caption{The crosses linked by a solid line show the abundance
(in Mpc$^{-3}$) of nearby galaxies with star-formation rates exceeding
a given value (in M$_\odot$/yr) plotted against that value. The data
are taken from Gallego et al (1995). The triangle is the abundance 
estimated at $z\sim 1.25$ from the data of Cowie et al (1995) while
the squares are similar estimates at $z\sim 3.25$ from the data of
Steidel et al (1996). Filled symbols assume $q_0=0.5$ while the open
square assumes $q_0=0.05$.}
\end{figure}

Cowie et al (1995) present redshift and emission line data 
for an almost complete sample of galaxies with $B<24.5$. Both
population modelling of the UV continuum and the emission line data 
in their figure 3 suggest that their sample should be complete for
objects with unobscured star-formation rates greater than 10
M$_\odot$/yr over the redshift range $1.0<z<1.5$. They find an average
of 0.7 such objects per square arcmin, leading to a mean comoving
abundance of $6.4\times 10^{-4}$Mpc$^{-3}$ assuming $q_0=0.5$. The 
same exercise can also be carried out at higher redshift. Steidel 
et al (1996) argue that their Lyman limit imaging technique finds 
all galaxies in the redshift range $3<z<3.5$ with $R>25$. This band 
corresponds to the far UV continuum in the galaxy rest frame, and 
they estimate $R>25$ to correspond to an unobscured 
star-formation rate exceeding 6.3 M$_\odot$/yr. They find 0.4 such objects 
per square arcmin giving a comoving volume density of $3.6\times 
10^{-4}$Mpc$^{-3}$. I plot both abundances on top of the local data in
figure 2. Remarkably, the abundance of systems forming stars faster 
than 10 M$_\odot$/yr is similar at $z=3.25$, at $z=1.25$ and at $z=0$. 
If one instead adopts $q_0=0.05$ the star-formation rate for
the Steidel et al galaxies increases by a factor of 3 but their
abundance drops by a factor of 5. As figure 2 shows the inferred
abundance is still similar to that seen locally. 

It thus appears that although global
star-formation rates were higher in the past, the
enhancement resulted purely from an increase in the number of systems
forming stars at modest rates. A new population of unobscured massive
starbursts is not seen at redshifts less than 3.5. It should, however,
be noted that the star-formation rates
are estimated in different ways for each of the three samples in figure
2; hence systematic shifts in the relative calibration might change
the impression given by this plot. 

\section{Millimeter observations of young galaxies}

In hierarchical models most star
formation occurs relatively late. (The equivalent of Figure 1
but for all the stars in the universe shows a much more gentle decline
to the present.) In addition there is no period when bulges and
ellipticals form rapidly as single units and so are very
bright. Instead star formation in the past occurred in lower mass and
more gas-rich objects than today but at rates comparable to
those in nearby systems. It seems unlikely that extreme objects like
the superluminous IRAS starburst sources were ever much more
abundant than they are today. The observational data in figure 2
support this conclusion but it is important to note that they refer
to unobscured star-formation. Most of the radiation in the strongest
nearby starbursts is absorbed by dust and reradiated in the far infrared.
In addition, the fraction of the energy emitted by dust seems to be
greater in systems which are more dynamically disturbed. Both
observational and theoretical arguments suggest that a larger
fraction of star-forming objects are irregular at high redshift, so
the fraction of the energy radiated in the FIR may be much
greater than it is locally. 

The amount of heated molecular gas and the amount of hot dust are the
key parameters which determine the observability of distant galaxies
by a millimeter array. A further important parameter is the size of the
emitting region. Recent studies of distant galaxies find that the
UV-emitting regions are almost always small with typical angular
sizes well below one arcsec, corresponding to physical sizes of
a kiloparsec or two. The physically-based models I have discussed make
significantly more pessimistic predictions than {\it ad hoc}
models which typic\-ally extrapolate the strong evolution detected in nearby
samples of IRAS galaxies in order to predict the FIR emission at $1<z<5$ (e.g. Rowan-Robinson, this volume). Objects with
star-formation rates of 30 M$_\odot$/yr are found in both the Cowie et al and Steidel
et al samples. Since these surveys covered very small areas it is likely that objects
will be found forming stars at 100 M$_\odot$/yr over the full redshift range $z<4$;
at least some of these objects may emit most of their radiation in the FIR and would
then be only a few times less luminous than nearby ultraluminous IR
galaxies (e.g. Clements et al 1996). Scaling from such systems
shows that an order of magnitude increase in sensitivity of millimeter
interferometers would allow line observations of young galaxies
at $z>1$ and detections of the dust emission from the most luminous
systems at high redshift ($z=4$ to 5). The detection of millimeter
lines will be particularly important because it will give
information about the dynamical state of the gas during the early
stages of galaxy assembly. An angular resolution approaching 0.1 arsec
will be needed to map the emitting gas in most systems.
%

%
%

\end{document}